\documentclass[11pt,a4paper]{article}
\usepackage{jheppub}

\usepackage{amsmath}
\usepackage{amsfonts}
\usepackage{amsthm}
\usepackage{slashed}
\usepackage{graphicx}
\usepackage{mathrsfs,bbm}
\usepackage{subfig}
\usepackage{booktabs}
\usepackage[numbers,sort&compress]{natbib}
\usepackage{xcolor}

\usepackage{multirow}
\usepackage{todonotes}

\usepackage{siunitx}
\makeatletter
\g@addto@macro\bfseries{\boldmath}
\makeatother

\theoremstyle{remark}
\newtheorem*{remark}{Remark}

\newcommand{\dimeps}{\varepsilon}
\newcommand{\td}{\mathrm{d}}
\newcommand{\iu}{\mathrm{i}}
\newcommand{\ieps}{\iu\epsilon}
\newcommand{\FI}{\mathcal{I}}
\newcommand{\TOP}[1][top]{\mathsf{#1}}
\newcommand{\PL}{\TOP[PL]}

\newcommand{\NA}{\TOP[NA]}
\newcommand{\NB}{\TOP[NB]}
\newcommand{\LO}{\TOP[LO]}
\newcommand{\FIPL}[1]{\FI_{\PL}(#1)}

\newcommand{\FINA}[1]{\FI_{\NA}(#1)}
\newcommand{\FINB}[1]{\FI_{\NB}(#1)}
\newcommand{\FILO}[1]{\FI_{\LO}(#1)}
\newcommand{\FIPLs}[1]{\FI_{\PL}^{(6)}(#1)}

\newcommand{\FINAs}[1]{\FI_{\NA}^{(6)}(#1)}
\newcommand{\FINBs}[1]{\FI_{\NB}^{(6)}(#1)}
\newcommand{\FILOs}[1]{\FI_{\LO}^{(6)}(#1)}
\newcommand{\F}{\mathcal{F}}
\newcommand{\U}{\mathcal{U}}

\newcommand{\hlog}[2]{G\left( #2 ; #1 \right)}
\newcommand{\HyperInt}{\href{http://bitbucket.org/PanzerErik/hyperint/}{\texttt{\textup{HyperInt}}}}

\title{Two-loop mixed QCD-EW corrections to $q \overline{q} \to H g$, $qg \to Hq$, and 
$\overline{q}g \to H\overline{q}$}

\author[a]{Marco Bonetti} 
\author[b]{Erik Panzer}
\author[c]{Lorenzo Tancredi}

\affiliation[a]{Institute for Theoretical Particle Physics and Cosmology, RWTH Aachen University, \\Sommerfeldstrasse 16, D-52056 Aachen, Germany} 
\affiliation[b]{Mathematical Institute, University of Oxford, \\OX2 6GG Oxford, UK}
\affiliation[c]{Physik Department, Technische Universität München, \\James-Franck-Straße 1, 85748 Garching, Germany}

\emailAdd{bonetti@physik.rwth-aachen.de}
\emailAdd{erik.panzer@maths.ox.ac.uk}
\emailAdd{lorenzo.tancredi@tum.de}

\abstract{
We compute the two-loop mixed QCD-Electroweak corrections to $q \overline{q} \to H g$ and its crossed channels $q g \to H q$, $\overline{q} g \to H \overline{q}$, limiting ourselves to the contribution of light virtual quarks. We compute the independent helicity amplitudes as well as the form factors for this process, expressing them in terms of hyperlogarithms with algebraic arguments. The Feynman integrals are computed by direct integration over Feynman parameters and the results are expressed in terms of a basis of rational prefactors.
}

\arxivnumber{2203.17202}

\preprint{\begin{minipage}[t]{8cm}\begin{flushright} Pre-print numbers\\
        TTK-22-11       \\
        P3H-22-025      \\
        TUM-HEP-1394/22
      \end{flushright}\end{minipage}}

\begin{document}

\maketitle

\catcode`\@=11
\font\manfnt=manfnt
\def\Watchout{\@ifnextchar [{\W@tchout}{\W@tchout[1]}}
\def\W@tchout[#1]{{\manfnt\@tempcnta#1\relax%
  \@whilenum\@tempcnta>\z@\do{%
    \char"7F\hskip 0.3em\advance\@tempcnta\m@ne}}}
\let\foo\W@tchout
\def\dubious{\@ifnextchar[{\@dubious}{\@dubious[1]}}
\let\enddubious\endlist
\def\@dubious[#1]{%
  \setbox\@tempboxa\hbox{\@W@tchout#1}
  \@tempdima\wd\@tempboxa
  \list{}{\leftmargin\@tempdima}\item[\hbox to 0pt{\hss\@W@tchout#1}]}
\def\@W@tchout#1{\W@tchout[#1]}
\catcode`\@=12


\section{Introduction}
\label{sec:intro}

The discovery of the Higgs boson at the CERN LHC in 2012~\cite{Aad:2012tfa,Chatrchyan:2012ufa} can be considered, beyond doubts, one of the greatest achievements of
the high-energy physics program of the 21st century. This long anticipated particle crowns the Standard Model as a solid theory of reality and the study of its properties 
has the potential to help elucidate the mechanism of Electroweak (EW) symmetry breaking and to shed light on possible new physics scenarios beyond the Standard Model.

Since the Higgs discovery, the LHC has demonstrated not only to be a discovery machine, but also an impressive precision physics laboratory. In fact, in the past decade 
astonishing experimental results have been obtained by the LHC collaborations, which have made it possible to rediscover the  Standard Model of particle physics
and to perform cutting edge precision measurements, in some cases way beyond the original expectations. 
As the result of these impressive performances, the theoretical particle physics community has taken over the challenge of pushing theoretical predictions to
similar degrees of precision.
In particular, while the exact experimental precision achievable depends on the process considered, it has been shown that reaching the goal of the 
$\sim \mathcal{O}(1\%)$ level accuracy at the LHC for various important observables is within reach~\cite{Dainese:2019rgk}. This will be possible not only
because of the large statistics that will be accumulated at the LHC and its high luminosity upgrade (HL-LHC), but also thanks
to refined techniques for the determination of interaction luminosity~\cite{ATLAS:2019pzw,CMS:2021xjt} and for jet reconstruction~\cite{CMS:2016lmd,ATLAS:2017bje}. 
Among the observables that are most promising, a special role is played by the 
Higgs transverse momentum ($p_\perp$)
distribution. The Higgs $p_\perp$ is particularly interesting since 
a precision comparison of the measured distribution with similarly precise theoretical
predictions could allow us, among other things, to obtain precious indirect information on the Higgs couplings and 
therefore on the details of the Spontaneous Symmetry Breaking (SSB) mechanism, 
see for example ref.~\cite{Bishara:2016jga}.

The theoretical determination of the Higgs $p_\perp$ distribution at the LHC at the percent precision
level involves the calculations of the production
of a Higgs boson in addition to QCD radiation in hadron-hadron scattering $pp \to H+X$,
which can be mediated at the elementary level by different types of interactions.
As it is well known, due to the very high gluon luminosity, 
the main channel of production for the Higgs boson at the LHC is 
 gluon fusion, in particular in those configurations where the Higgs boson couples to 
gluons through a loop of top quarks. 
This interaction process has been computed to Next-to-Next-to-Leading-Order (NNLO) in pure QCD 
retaining full dependence on the top-quark mass~\cite{Georgi:1977gs,Graudenz:1992pv,Spira:1995rr,Aglietti:2006tp,Czakon:2020vql}, 
while it is known up to $\textup{N}^3\textup{LO}$ in 
QCD in the limit of infinite top-quark mass~\cite{Anastasiou:2015vya,Mistlberger:2018etf}. 
This so-called Higgs Effective Field Theory (HEFT) has proven to be surprisingly reliable to describe higher-order QCD contributions. In fact, in this approximation,
the  $\textup{N}^3\textup{LO}$ amounts to around $5\%$ of the total cross section with a scale uncertainty of $\sim 2\%$~\cite{Anastasiou:2016cez}.
At this impressive level of precision, a proper treatment of sub-leading production modes and uncertainties becomes mandatory. 
Together with the absence of $\textup{N}^3\textup{LO}$ parton distribution functions, 
the two other main sources of theoretical uncertainty
are the impact of finite quark-mass effects and the NLO QCD corrections to the electroweak (EW) production modes~\cite{Anastasiou:2016cez}, which usually go under the name of QCD-EW corrections
and are the main focus of this paper. 

As is the case in pure QCD, also QCD-EW corrections can affect
different production channels.
The most important 
one at the LHC is once again the gluon fusion channel.
QCD-EW corrections to this production channel contribute in two different configurations: 
the first class of diagrams involves the effect of 
EW corrections to the standard
$ggH$ coupling through a massive top loop. This contribution 
has been shown to be extremely small already at LO, 
contributing to less than $1\%$ to the total cross-section~\cite{Degrassi:2004mx}.
The second class of diagrams involves instead a loop of light quarks which connects the gluons to a pair of EW bosons, which then fuse to generate the Higgs.
This production channel increases the total cross section at LO by around $5\%$ with respect 
to the pure QCD corrections~\cite{Aglietti:2004nj,Aglietti:2004tq,Aglietti:2006tp,Aglietti:2007as,Bonciani:2003cj,Bonciani:2003te,Bonciani:2008az,Bonciani:2009nb,Bonetti:2016brm} and,
since NLO QCD corrections to Higgs production in gluon fusion 
can be as large as the LO contribution, a full account of these higher order corrections
is essential to keep the theory uncertainty 
under control at the percent level.
NLO QCD-EW effects in this channel have been estimated first in the unphysical 
limit $m_H \ll m_V$ with $V=W,Z$ in ref.~\cite{Anastasiou:2008tj}, 
then by treating the real emissions in the soft-gluon approximation in ref.~\cite{Bonetti:2017ovy,Bonetti:2018ukf} and more recently with full dependence on the Higgs and the vector boson masses and on the external QCD radiation in ref.~\cite{Becchetti:2018xsk,Bonetti:2020hqh,Becchetti:2020wof,Becchetti:2021axs}. 

In addition to the gluon-initiated diagrams, mixed QCD-EW corrections to $pp \to H+X$ 
 also receive
contributions from diagrams that involve a pair of quarks.
There are in particular three relevant partonic channels, $qg \to Hq$, $\overline{q}g \to H\overline{q}$, and $q\overline{q} \to Hg$, which at LO proceed through a 
loop of virtual electroweak 
vector bosons. Their impact both at Tevatron and at the LHC was estimated for the first time in~\cite{Keung:2009bs} 
for a Higgs mass of $m_H = \SI{120}{\giga\eV}$. 
There it was shown that, as expected,
their relative weight at the LHC is smaller than at the Tevatron, due to the much larger gluon luminosity, but it is still sizable: in particular, the LO diagrams are responsible for a $- 3 \%$
shift in the $p_\perp$ distribution for values $\SI{100}{\giga\eV}\leq p_\perp \leq \SI{300}{\giga\eV}$.
With the goal of pushing theoretical uncertainties in Higgs distributions at the percent level, 
it is therefore important to consider also NLO corrections to this channel, first
of all to obtain a reliable estimate of their uncertainty, and also to account for their possible 
non-trivial effects on the shape of the distributions.

One-loop mixed QCD-EW contributions containing light quarks as external states have been thoroughly investigated in \cite{Hirschi:2019fkz}, therefore the last missing building blocks are the relevant two-loop amplitudes.
In this paper, we focus on their analytic calculation,
liming ourselves to those contributions which involve only massless virtual quarks, similarly to what was done in \cite{Bonetti:2020hqh}. 
Together with the phenomenological motivation, 
this calculation is interesting also from a formal point of view. Indeed,
despite some similarities with the purely gluonic case described in \cite{Bonetti:2020hqh}, 
the quark induced channels are substantially more complicated for at least two reasons. First of all, they involve new classes of integrals that were entirely absent in the gluon induced ones. Moreover, at variance with the gluonic channel, 
these amplitudes have a non-trivial infrared structure and, for this reason, their finite remainders require a larger number of weight four functions.
From a purely mathematical  viewpoint, this calculation is particularly interesting due to
the presence of a large number of square roots in the symbol of the relevant 
two-loop master integrals. In fact, while all integrals turn out to be linearly reducible~\cite{Brown:2008um,Brown:PeriodsFeynmanIntegrals} and can therefore be expressed in terms of hyperlogarithms
by direct integration over their Feynman-Schwinger parameter representation
algorithmically~\cite{Panzer:2014caa}, the presence of a large number of
square roots renders the application of standard algorithms to handle and simplify
the resulting expressions highly non-trivial.

The paper is organized as follows. In \autoref{sec:notation} we discuss the structure of the amplitude, and in particular the treatment of vector and axial EW couplings; we then describe how to relate the helicity amplitudes to the two form factors composing the amplitude, and how to extract said form factors; at last, we give our convention for the integral families and integration measure. In \autoref{sec:compt} we discuss the computational framework we used, first to evaluate the master integrals, and secondly to reduce the expressions to a minimal set of algebraic prefactors times a set of polylogarithmic functions. In \autoref{sec:Cat} we discuss the UV renormalization of the amplitude and the structure of IR divergences, extracting the finite remainder for the virtual corrections. We review our results and draw our conclusions in \autoref{conc}.

We provide the list of master integrals for the LO and virtual NLO amplitude in \autoref{sec:masters}. The analytic expressions of the finite remainders, as well as of the master integrals, are provided in the supplementary material to this paper.


\section{The scattering amplitude}
\label{sec:notation}

\begin{figure}[ht]
\centering
	\subfloat[]{{\includegraphics[width=0.29\textwidth]{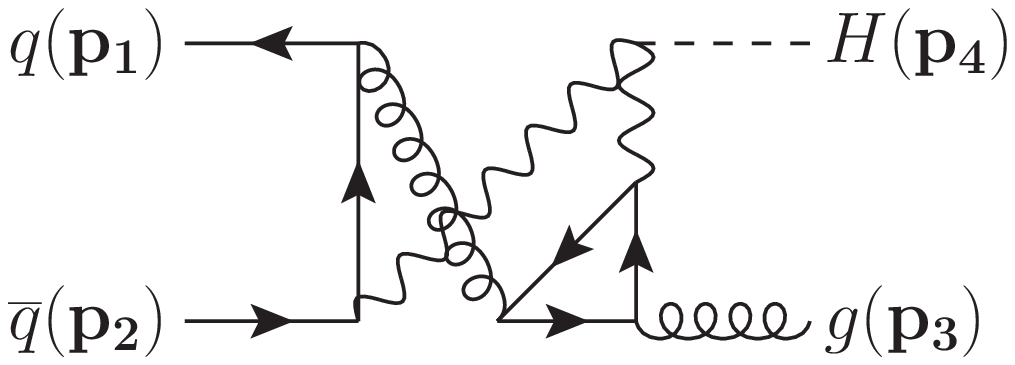}}\label{fig:diagrams3}}
	\qquad
	\subfloat[]{{\includegraphics[width=0.29\textwidth]{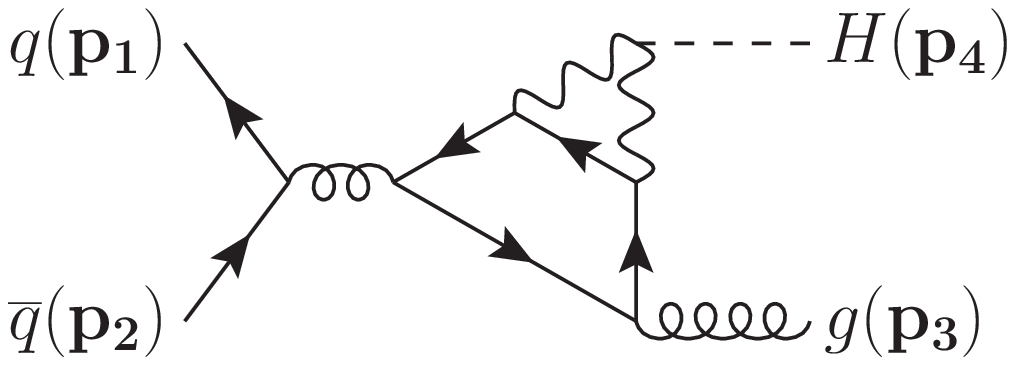}}\label{fig:diagrams4}}
	\qquad
	\subfloat[]{{\includegraphics[width=0.29\textwidth]{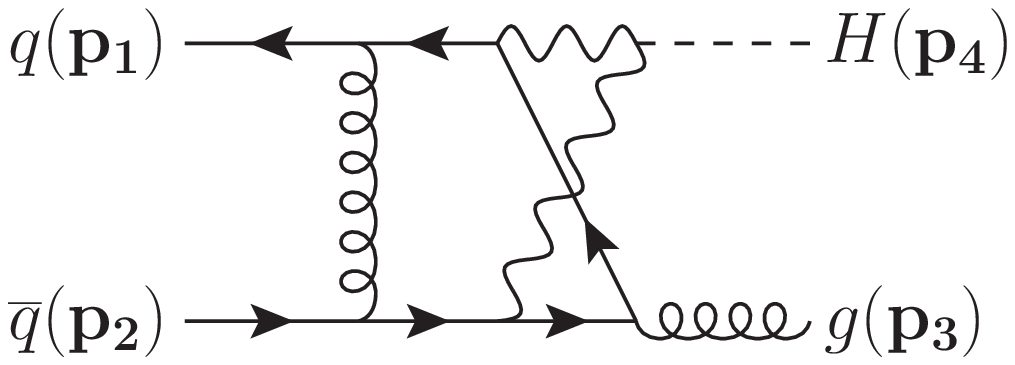}}\label{fig:diagrams1}}
	\caption{Representative diagrams for the process $H \to q \overline{q} g$. The internal wavy lines represent massive vector bosons. All momenta are taken to be incoming.}
    \label{fig:diagrams}
\end{figure}

Our goal is to compute the two-loop mixed QCD-EW corrections 
to the three partonic scattering processes $q g \to H q$, $\overline{q} g \to H \overline{q}$, 
and $q \overline{q} \to H g$. To do so, we start by considering the decay of a Higgs
boson to a quark-antiquark pair and a gluon
\begin{equation}
	H(\mathbf{p}_4) \to q(\mathbf{p}_1) + \overline{q}(\mathbf{p}_2) + g(\mathbf{p}_3),
\end{equation}
where the Higgs boson couples to the quarks through a pair of massive vector bosons $V$, where $V$ is either equal to $W^\pm$ or $Z$, see \autoref{fig:diagrams}.

The scattering amplitude for this process, $\mathcal{M}$, depends on the three Mandelstam variables
\begin{align}
s = (p_1 + p_2)^2\,,
\quad
t = (p_1 + p_3)^2\,,
\quad
u = (p_2 + p_3)^2\,,
\quad
\textup{with} \quad s+t+u = m_h^2\,,
\end{align}
and on the mass of the vector boson that mediates the interaction between the Higgs boson and the massless quarks, denoted as $m_V$. Throughout, $m_h$ indicates the Higgs boson mass.
The dependence of the scattering amplitude on the $\mathrm{SU}(3)$ color structure is given by the Gell-Mann matrices 
$T^{c_3}_{i_1 i_2}$, where $c_3$ is the color index associated with the gluon, and $i_1$ ($i_2$) is the color index of the quark (antiquark)
\begin{align}
\mathcal{M}_{s_1 s_2 \lambda_3}(\mathbf{p}_1,\mathbf{p}_2,\mathbf{p}_3) &= 
	\left[ \frac{\alpha^{3/2} m_W}{2 \sin^3 \theta_W} \right] T^{c_3}_{i_1 i_2}  A_{s_1 s_2 \lambda_3}(\mathbf{p}_1,\mathbf{p}_2,\mathbf{p}_3) \nonumber \\ &=  
	\left[	\frac{\alpha^{3/2} m_W}{2 \sin^3 \theta_W} \right] T^{c_3}_{i_1 i_2}  \epsilon^{*\mu}_{\lambda_3}(\mathbf{p}_3) \overline{u}_{s_1}(\mathbf{p}_1) \mathcal{A}_\mu(s,t,u,m_V^2) v_{s_2}(\mathbf{p}_2). \label{eq:ampl}
\end{align}
In Eq.~\eqref{eq:ampl} we have collected out the overall electroweak
coupling and we also made explicit the dependence on the spin of the quarks 
(${s_1,s_2}$) and on the
polarization vector of the gluon $\epsilon_{\lambda_3}$
which satisfies $\epsilon_{\lambda_3} \cdot p_3 = 0$.
In addition, Ward Identities require that the amplitude $\mathcal{A}_\mu(s,t,u,m_V^2)$ must satisfies the transversality condition
\begin{align}
    \label{polcon}
    p_3 \cdot \mathcal{A}(s,t,u,m_V^2) = 0 \,.
\end{align}
We write the coupling of the vector boson $V$ with the light quarks as $g_V^v + \gamma_5 g_V^a$, where~\cite{Romao:2012pq}
\begin{align}\label{eq:couplings}
    \begin{aligned}
        g_W^v &= -\iu \frac{e}{\sin \theta_W}\frac{1}{2\sqrt{2}}\,, \quad &g_W^a &= +\iu \frac{e}{\sin \theta_W}\frac{1}{2\sqrt{2}}\,, \\
        g_{Zf}^v &= -\iu \frac{e}{\sin \theta_W \cos \theta_W} \left[\frac{T_f}{2} - Q_f \sin^2 \theta_W\right] \,, \quad &g_{Zf}^a &= -\iu \frac{e}{\sin \theta_W \cos \theta_W} \left[\frac{T_f}{2}\right]\,.
    \end{aligned}
\end{align}
$Q_f$ and $T_f$ are the electric charge and the eigenvalue of the third generator of $\mathrm{SU}(2)_L$, both associated to the fermion $f$ interacting with $V$, $e$ is the absolute value of the electric charge of the electron, and $\theta_W$ is the weak mixing angle. In what follows we also define $e = \sqrt{4\mathrm{\pi}\alpha}$ for the electroweak coupling and $g_S = \sqrt{4\mathrm{\pi}\alpha_S}$ for the strong coupling.

It is useful to separate the amplitude into different contributions, depending on how the electroweak bosons are coupled to the rest of the diagram; clearly, depending on the number of loops we are interested in, different contributions can play a role. If we limit ourselves to consider Feynman diagrams up to order $\mathcal{O}( \alpha^{3/2} \alpha_S^{3/2} )$,
there are only two possible non-zero classes of diagrams that can contribute, 
which we call \emph{open} and \emph{closed}. In particular, we write
\begin{equation}
\mathcal{A}_{\mu}(\mathbf{p}_1,\mathbf{p}_2,\mathbf{p}_3) = 
\mathcal{A}^{\textup{open}}_{\mu}(\mathbf{p}_1,\mathbf{p}_2,\mathbf{p}_3)
+ \mathcal{A}^{\textup{closed}}_{\mu}(\mathbf{p}_1,\mathbf{p}_2,\mathbf{p}_3)
\end{equation}
where $\mathcal{A}_\mu^{\textup{open}}$ receives contribution from those diagrams where the two vector bosons are both attached to the 
external fermion line, see Figure \autoref{fig:diagrams1}, while $\mathcal{A}_\mu^{\textup{closed}}$ encompasses the diagrams where both vector bosons couple to a closed fermion loop, see Figure \autoref{fig:diagrams4}.

Let us start by considering  $\mathcal{A}_\mu^{\textup{open}}$. Feynman diagrams of this type start contributing at one-loop order. The EW bosons couple to the fermion line both through vector and axial terms (depending on the chirality of the external quarks) but it is possible to bypass the explicit manipulation of terms containing $\gamma_5$ by considering polarized external states: by anticommuting $\gamma_5$ until it touches the spinor $\overline{u}_{L/R}(\mathbf{p}_1)$, such a matrix coming from the EW vertices is absorbed into the left- or right-chirality projectors $\mathbb{P}_{L} = \frac{1-\gamma_5}{2}$ and $\mathbb{P}_{R} = \frac{1+\gamma_5}{2}$, respectively. 
The resulting expressions for the helicity amplitudes do not contain $\gamma_5$ anymore and can be computed 
assuming that the electroweak bosons only couple through a vector current to the fermion line,
with an overall coupling coefficient determined by type and chirality of the quark $q(\mathbf{p}_1)$. Specifically, for $\overline{u}_{L}(\mathbf{p}_1) = \overline{u}(\mathbf{p}_1) \mathbb{P}_L$ we obtain
\begin{align}
	\label{AopenP}
    \begin{aligned}
		\mathcal{A}^{\textup{open}}_{L,\mu} = \left( \frac{2}{\cos^4 \theta_W} Q_q^2 \sin^4 \theta_W \right) \mathbb{P}_L \left[ \tau_{1,\mu} F_{1,m_Z}^{\textup{open}} + \tau_{2,\mu} F_{2,m_Z}^{\textup{open}} \right]	, 
	\end{aligned}
\end{align}
where, as in eq.~\eqref{eq:couplings}, $Q_q$ is the electric charge of the quark $q(\mathbf{p}_1)$, and we define
\begin{align}
    F_{i,m_X} = F_i(s,t,u,m_h^2,m_V^2 = m_X^2)  \,,
\end{align}
and
\begin{align}
    \tau_{1\mu} = \slashed{p}_3 p_{2\mu} - p_2 \cdot p_3 \gamma_\mu  \,,\qquad\qquad\tau_{2,\mu} = \slashed{p}_3 p_{1\mu} - p_1 \cdot p_3 \gamma_\mu \,. \label{eq:taus}
\end{align}
For $\overline{u}_{R}(\mathbf{p}_1) = \overline{u}(\mathbf{p}_1) \mathbb{P}_R$ we get
\begin{align}
	\label{AopenM}
	\begin{aligned}
		\mathcal{A}^{\textup{open}}_{R,\mu} &=  \left\{\mathbb{P}_R \left[ \tau_{1,\mu} F_{1,m_W}^{\textup{open}} + \tau_{2,\mu} F_{2,m_W}^{\textup{open}} \right]  
		\right.\\&~~~~~~~\left.
		+ \frac{2}{\cos^4 \theta_W}\left( T_q - Q_q \sin^2 \theta_W \right)^2 \mathbb{P}_R \left[ \tau_{1,\mu} F_{1,m_Z}^{\textup{open}} + \tau_{2,\mu} F_{2,m_Z}^{\textup{open}} \right]\right\}	.
	\end{aligned}
\end{align}
The different form of the couplings between Eq.~\eqref{AopenP} and Eq.~\eqref{AopenM} is due to the fact that $\mathcal{A}^{\textup{open}}_{L,\mu}$ receives contributions from the $Z$ boson only, while $\mathcal{A}^{\textup{open}}_{R,\mu}$ is sensitive to the $W$ boson as well. We assume the Cabibbo--Kobayashi--Maskawa mixing matrix to be the identity matrix.\footnote{This approximation is justified because the off-diagonal entries are strongly suppressed.}

~

Starting at two loops, one can draw diagrams where either one or both EW bosons couple to an internal light-quark loop, see 
Figures~(\ref{fig:diagrams3}) and (\ref{fig:diagrams4}), respectively. For those diagrams where a single EW boson attaches to the fermion loop, it is possible to show that both the vector and the axial contribution vanish at the level of the amplitude: the axial part cancels out adding together degenerate isospin doublets, while the vector contribution must add up to zero due 
to Furry's theorem. In case of both the EW vector bosons attached to the same light-quark loop, two contributions are possible: one proportional to $(g_V^v)^2 + (g_V^a)^2$ and independent from $\gamma_5$, and one proportional to $g_V^v g_V^a \gamma_5$. This last contribution vanishes identically when summing over degenerate isospin doublets. This clearly does not apply to the third quark doublet. In diagrams containing $W^\pm$ bosons we avoid the issue by not allowing top and bottom quarks circulating in the internal loop, while we consider all but the top quark in loops coupled with $Z$ bosons, since we expect that the missing axial contributions will be suppressed. 

We collect the contribution of the diagrams containing a closed quark loop to the amplitude $\mathcal{A}_\mu$ under the label $\mathcal{A}^{\textup{closed}}_\mu$. As explained above, up to their contribution to the overall coupling factor, 
the amplitude contains no axial terms and it can be decomposed in form factors 
following the same procedure as for the amplitude $H \to q \overline{q} g$ in pure QCD, see \cite{Melnikov:2017pgf}. 
It reads
\begin{align}
	\label{Aclosed}
	\begin{aligned}
		\mathcal{A}^{\textup{closed}}_\mu & = \frac{1}{2} \left\{ 4 \left[ \tau_{1,\mu} F_{1,m_W}^{\textup{closed}} + \tau_{2,\mu} F_{2,m_W}^{\textup{closed}} \right]    \right.\\
		&\left. + \frac{2}{\cos^4 \theta_W}\left( \frac{5}{4}-\frac{7}{3}\sin^2 \theta_W+\frac{22}{9}\sin^4 \theta_W \right) \left[ \tau_{1,\mu} F_{1,m_Z}^{\textup{closed}} + \tau_{2,\mu} F_{2,m_Z}^{\textup{closed}} \right] \right\}	,
	\end{aligned}
\end{align}
where the factor of $4$ in the first line comes from diagrams containing the $W$ boson (summed over the first two generations of quarks),
while the term in round brackets in the second line stems from diagrams featuring the $Z$ boson (summing on all quarks except the top). We stress also that, since the
contribution $\mathcal{A}^{\textup{closed}}_\mu$ only starts at the two-loop order, we have collected
out one more factor of $\alpha_S^{1/2}$, compared to $\mathcal{A}^{\textup{open}}_\mu$.

The form factors $F_{j,m_V}^{\textup{open}}$ and $F_{j,m_V}^{\textup{closed}}$ that contribute to Eqs.~\eqref{AopenP}, \eqref{AopenM}, and \eqref{Aclosed} admit an expansion in the strong coupling constant
\begin{align}
    \begin{aligned}
        F_{j,m_V}^{\textup{open}} &= \sqrt{\alpha_S} \left[F_{j,m_V}^{\textup{open},(1)} + \left( \frac{\alpha_S}{2 \mathrm{\pi}} \right) F_{j,m_V}^{\textup{open},(2)} + \mathcal{O}\left(\alpha_S^2 \right)\right] , \qquad \\
        F_{j,m_V}^{\textup{closed}} &= \sqrt{\alpha_S} \left[
        \left( \frac{\alpha_S}{2 \mathrm{\pi}} \right) F_{j,m_V}^{\textup{closed},(2)} + \mathcal{O}\left( \alpha_S^2 \right)\right] ,
    \end{aligned}
\end{align}
where $F_{j,m_V}^{\textup{class},(l)}$ is the $l$-loop contribution of the corresponding class
of diagrams and, as always, $m_V$ can be either $m_Z$ or $m_W$. 
Finally, while $F_{j,m_V}^{\textup{open},(1)}$ and $F_{j,m_V}^{\textup{closed},(2)}$ have a trivial color structure, $F_{j,m_V}^{\textup{open},(2)}$ can be decomposed as
\begin{align}
    F_{j,m_V}^{\textup{open},(2)} = N_c F_{j,m_V}^{N_c} + \frac{1}{N_c} F_{j,m_V}^{1/N_c}	\,,
\end{align}
where $N_c$ is the number of colors.

\subsection{Helicity amplitudes}

When dealing with scattering amplitudes containing massless external states, 
helicity amplitudes are often the simplest physical objects to compute.
We fix the helicities of the external quark-antiquark pair and of the external
gluons by use of the spinor-helicity formalism
\begin{align}
	\begin{aligned}
		\mathbb{P}_R u(\mathbf{p}) = u_R(\mathbf{p}) = v_R(\mathbf{p}) = |p\rangle \,,	\qquad
		&\mathbb{P}_L u(\mathbf{p}) = u_L(\mathbf{p}) = v_L(\mathbf{p}) = |p] \,,	\\
		\overline{u}(\mathbf{p}) \mathbb{P}_L = \overline{u}_L(\mathbf{p}) = \overline{v}_L(\mathbf{p}) = [p| \,,	\qquad
		&\overline{u}(\mathbf{p}) \mathbb{P}_R = \overline{u}_R(\mathbf{p}) = \overline{v}_R(\mathbf{p}) = \langle p| \,,	\\
		\epsilon^{*\mu}_+(\mathbf{p}_3) = + \frac{\langle q \gamma^\mu 3]}{\sqrt{2} \langle q 3 \rangle},	\qquad
		&\epsilon^{*\mu}_-(\mathbf{p}_3) = - \frac{[q \gamma^\mu 3\rangle}{\sqrt{2} [q 3]}	\,,
	\end{aligned}
\end{align}
where $q$ is a generic light-like vector that represents the gauge freedom associated
to the external gluon. As in Eq.~\eqref{eq:ampl}, we indicate the helicity amplitudes as $A_{s_1 s_2 \lambda_3}$, where $s_1,s_2$ are the spins of the quarks while $\lambda_3$ is the polarisation of the gluon.
There are two independent helicity amplitudes, which we choose to be $A_{RL +}$ and $A_{LR +}$,
while all other helicity configurations can be obtained from these two by parity and charge
conjugation transformations, as exemplified below.

In the same way as we did for the full amplitude $\mathcal{A}_\mu$, we  
divide the helicity amplitude for $H \to q \overline{q} g$ into two separately
gauge-invariant contributions
\begin{align}
	A_{s_1 s_2 \lambda_3} = A^{\textup{open}}_{s_1 s_2 \lambda_3} + A^{\textup{closed}}_{s_1 s_2 \lambda_3} \,.
\end{align}
By explicitly fixing the helicities of the external particles and keeping track of the
electroweak couplings we find for the two different contributions
\begin{align}
	A^{\textup{open}}_{RL+} &= \left[ F_{1,m_W}^{\textup{open}} + \frac{2}{\cos^4 \theta_W}\left( T_q - Q_q \sin^2 \theta_W \right)^2 F_{1,m_Z}^{\textup{open}} \right] \frac{s}{\sqrt{2}}\frac{[23]^2}{[12]}	\,, \\[1em]
	A^{\textup{open}}_{LR+} &= \left[ \frac{2}{\cos^4 \theta_W} Q_q^2 \sin^4 \theta_W F_{2,m_Z}^{\textup{open}} \right] \frac{s}{\sqrt{2}}\frac{[13]^2}{[12]}
\end{align}
and
\begin{align}
	A^{\textup{closed}}_{RL+} =  &\frac{1}{2} \left[ 4 F_{1,m_W}^{\textup{closed}} 
	+ \frac{2}{\cos^4 \theta_W}\left( \frac{5}{4}-\frac{7}{3}\sin^2 \theta_W+\frac{22}{9}\sin^4 \theta_W \right) F_{1,m_Z}^{\textup{closed}} \right] \frac{s}{\sqrt{2}}\frac{[23]^2}{[12]}	\,,\\[1em]
	A^{\textup{closed}}_{LR+} =  &\frac{1}{2} \left[ 4 F_{2,m_W}^{\textup{closed}} 
	+ \frac{2}{\cos^4 \theta_W}\left( \frac{5}{4}-\frac{7}{3}\sin^2 \theta_W+\frac{22}{9}\sin^4 \theta_W \right) F_{2,m_Z}^{\textup{closed}} \right] \frac{s}{\sqrt{2}}\frac{[13]^2}{[12]}	\,.
\end{align}
We stress that for $A^{\textup{open}}_{s_1 s_2 \lambda_3}$ special care has to be taken to derive these formulas, taking into account the parity violating coupling of the EW bosons to the external quark line.

The remaining non-zero helicity amplitudes can be obtained by parity and charge conjugation
transformations. For the \emph{closed} contributions we get
\begin{align}
    A^{\textup{closed}}_{RL-}(s,t,u,m_V^2) &=
    \left( A^{\textup{closed}}_{LR+}(s,t,u,m_V^2) \right)^*   \,, \label{eq:helclosed1}\\
        A^{\textup{closed}}_{LR-}(s,t,u,m_V^2) &=
    \left( A^{\textup{closed}}_{RL+}(s,t,u,m_V^2) \right)^*   \,, \label{eq:helclosed2}
\end{align}
where complex conjugation here is intended to act only on the spinor products: $[ ij ]^* = \langle ji \rangle$ and $\langle ij \rangle^* = [ ji ]$.

The \emph{open} contributions require a bit more care to be obtained, and explicitly read
\begin{align}
    A^{\textup{open}}_{RL-} &= \left[ F_{2,m_W}^{\textup{open}} + \frac{2}{\cos^4 \theta_W}\left( T_q - Q_q \sin^2 \theta_W \right)^2 F_{2,m_Z}^{\textup{open}} \right] \frac{s}{\sqrt{2}}\frac{\langle13\rangle^2}{\langle12\rangle} 	\,, \label{eq:helopen1}\\
    A^{\textup{open}}_{LR-} &= \left[ \frac{2}{\cos^4 \theta_W} Q_q^2 \sin^4 \theta_W F_{1,m_Z}^{\textup{open}} \right] \frac{s}{\sqrt{2}}\frac{\langle23\rangle^2}{\langle12\rangle}     \,.\label{eq:helopen2}
\end{align}

It is important to notice that, from their very definition in Eqs.~\eqref{AopenM} and \eqref{Aclosed}, the two form factors $F_1^{\textup{closed}}$ and $F_2^{\textup{closed}}$ are not independent, and instead are mapped into each other by exchanging $p_1$ and $p_2$, namely
\begin{align}
    \label{swap1}
    F_1^{\textup{closed}}(s,t,u,m_V^2) = F_2^{\textup{closed}}(s,u,t,m_V^2) \,,
\end{align}
implying that
\begin{align}
    \label{swapAclosed}
    \begin{aligned}
        A^{\textup{closed}}_{RL+}(s,t,u,m_V^2) &= A^{\textup{closed}}_{LR+}(s,u,t,m_V^2)    \,,\\
        A^{\textup{closed}}_{RL-}(s,t,u,m_V^2) &= A^{\textup{closed}}_{LR-}(s,u,t,m_V^2)    \,.
    \end{aligned}
\end{align}
Similarly, we observe that
\begin{align}
    \label{swap2}
    F_1^{\textup{open}}(s,t,u,m_V^2) = F_2^{\textup{open}}(s,u,t,m_V^2) \,,
\end{align}
while an identity analogous to Eq.~\eqref{swapAclosed} does not hold in this case due to the dependence of the EW coupling on the helicity of the external particles.

We checked our expressions for the helicity amplitudes at LO against the computer code \texttt{OpenLoops} \cite{Buccioni:2019sur}, finding excellent agreement.

\subsection{Form factor evaluation}
\label{sec:ff-evaluation}

Here we focus on the calculation of the two-loop corrections to
the form factors described above, namely to $F_{j,m_V}^{\textup{class}}$
with $V=W^\pm,Z$, and for the two classes of open and closed diagrams.
At the two-loop order, the form factors for a given type of EW boson $V$
receive contribution from $45$ different two-loop Feynman diagrams, see \autoref{fig:diagrams}.
We generate them with the program \texttt{qgraf} \cite{Nogueira:1991ex}.
The form factors $F_j^{\textup{class}}$ can then be extracted by applying the projectors $\mathcal{P}_{j}$ to the Feynman diagrams that contribute to $A^{\textup{class}}_{s_1 s_2 \lambda_3}$, eq.~\eqref{eq:ampl}, 
such as
\begin{align}
	F_j^{\textup{class}} = \sum_{s_1, s_2, \lambda_3} A^{\textup{class}}_{s_1 s_2 \lambda_3}(\mathbf{p}_1,\mathbf{p}_2,\mathbf{p}_3) \mathcal{P}_{j,\lambda_3 s_2 s_1}	\,,
\end{align}
where 
\begin{align}
	\mathcal{P}_{1,\lambda_3 s_2 s_1} &= \epsilon_{\lambda_3}^\nu(\mathbf{p}_3) \frac{1}{2(D-3)s t}\left[ \frac{D-2}{t}T_{1,\nu s_2 s_1}^\dagger - \frac{D-4}{u}T_{2,\nu s_2 s_1}^\dagger \right]	,\\
	\mathcal{P}_{2,\lambda_3 s_2 s_1} &= \epsilon_{\lambda_3}^\nu(\mathbf{p}_3) \frac{1}{2(D-3)s u}\left[ \frac{D-2}{u}T_{2,\nu s_2 s_1}^\dagger - \frac{D-4}{t}T_{1,\nu s_2 s_1}^\dagger \right]	,
\end{align}
with $T_{i,\mu s_1 s_2} = \overline{u}_{s_1}(\mathbf{p}_1) \tau_{i,\mu} v_{s_2}(\mathbf{p}_2) $
and $ \tau_{i,\mu}$ as defined in Eq.~\eqref{eq:taus}.

We use \texttt{FORM} \cite{Ruijl:2017dtg,Kuipers:2013pba} to apply the projectors on the individual Feynman diagrams,  carry out the Dirac algebra and express the form factors as linear combinations of scalar two-loop Feynman integrals. To describe all the integrals appearing in the amplitude we use three different integral families: one for planar integrals and two for non-planar integrals. We define such families as follows:
\begin{align}
	\label{eq:FI}
	\FI_{\TOP}(a_1,a_2,\ldots,a_8,a_9) =
	\int \frac{\mathfrak{D}^d k_1 \,\mathfrak{D}^d k_2}{D_1^{a_1} D_2^{a_2}  D_3^{a_3}  D_4^{a_4}  D_5^{a_5}  D_6^{a_6}  D_7^{a_7}  D_8^{a_8}  D_9^{a_9}}	\,,
\end{align}
where $\TOP \in \{\PL,\NA,\NB\}$ labels the families and the denominators $D_1,\ldots,D_9$ are given in \autoref{tab:topos}.
We use dimensional regularization with $d=4-2\dimeps$, and our convention for the integration measure for each loop is
\begin{align}
	\label{eq:loop-measure}
	\mathfrak{D}^d k = \frac{\td^d k}{\iu \pi^{d/2} \Gamma(1+\dimeps)}	\,.
\end{align}

\begin{table}[t]
	\small
	\centering
	\begin{tabular}{cl@{\qquad}l@{\qquad}l}
\toprule
Denominator	&Integral family $\PL$			&Integral family $\NA$		&Integral family $\NB$ 	\\
\midrule
$D_1$		&$k_1^2$						&$k_1^2$					&$k_1^2$					\\
$D_2$		&$k_2^2-m_V^2$					&$(k_1-k_2)^2$				&$(k_1-k_2)^2$				\\
$D_3$		&$(k_1-k_2)^2$					&$(k_1-p_1)^2$				&$(k_1-p_1)^2$				\\
$D_4$		&$(k_1-p_1)^2$					&$(k_2+p_3)^2-m_V^2$		&$(k_2+p_3)^2-m_V^2$		\\
$D_5$		&$(k_1-p_1-p_2)^2$				&$(k_1-p_1-p_2)^2$			&$(k_1-p_1-p_2)^2$			\\
$D_6$		&$(k_1-p_1-p_2-p_3)^2$			&$(k_2-p_1-p_2)^2-m_V^2$	&$(k_2-p_1-p_2)^2-m_V^2$	\\
$D_7$		&$(k_2-p_1-p_2-p_3)^2-m_V^2$	&$(k_1-k_2-p_3)^2$			&$(k_1-k_2-p_3)^2$			\\
$D_8$		&$(k_2-p_1)^2$					&$(k_2-p_1)^2-m_V^2$		&$(k_2-p_1)^2$				\\
$D_9$		&$(k_2-p_1-p_2)^2$				&$(k_1-p_1-p_3)^2$			&$(k_1-p_1-p_3)^2$			\\
\bottomrule
	\end{tabular}
	\caption{Definition of the planar ($\PL$) and non-planar ($\NA$ and $\NB$) integral families. The loop momenta are denoted by $k_1$ and $k_2$, while $m_V$ indicates the mass of the vector boson. The prescription $+ \ieps$ is understood for each propagator and not written explicitly.}
	\label{tab:topos}
\end{table}

\begin{figure}[ht]
\centering
	\subfloat[T1: $\FINB{1,1,1,1,0,1,1,1,0}$]{{\includegraphics[width=0.30\textwidth]{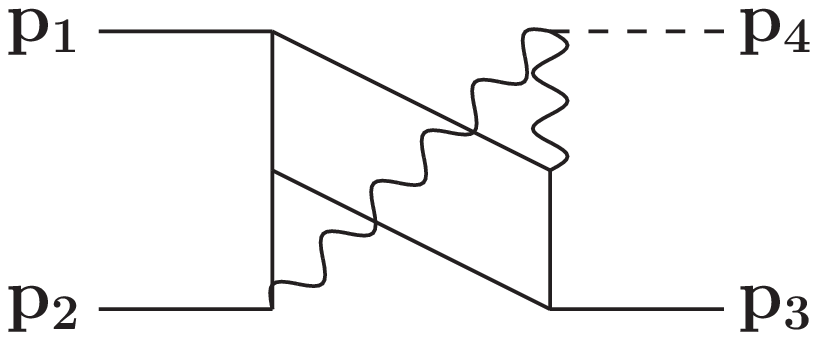}}\label{fig:T1}}
	\qquad
	\subfloat[T2: $\FINA{1,1,1,1,1,1,1,0,0}$]{{\includegraphics[width=0.30\textwidth]{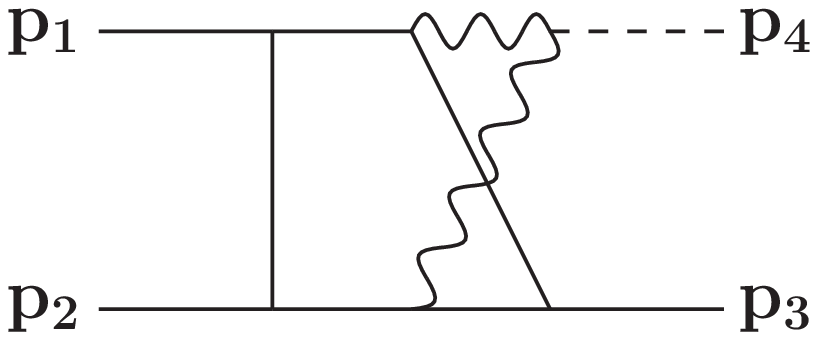}}\label{fig:T2}}
	\\
	\subfloat[T3: $\FIPL{0,1,1,1,1,0,1,1,1}$]{{\includegraphics[width=0.30\textwidth]{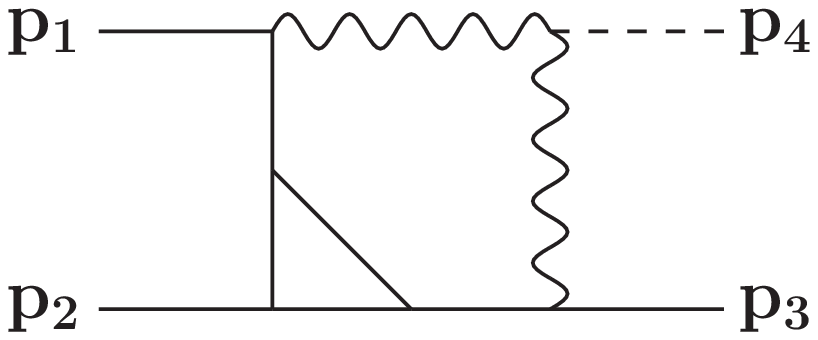}}\label{fig:T3}}
	\qquad
	\subfloat[T4: $\FIPL{1,1,1,1,0,0,1,1,1}$]{{\includegraphics[width=0.30\textwidth]{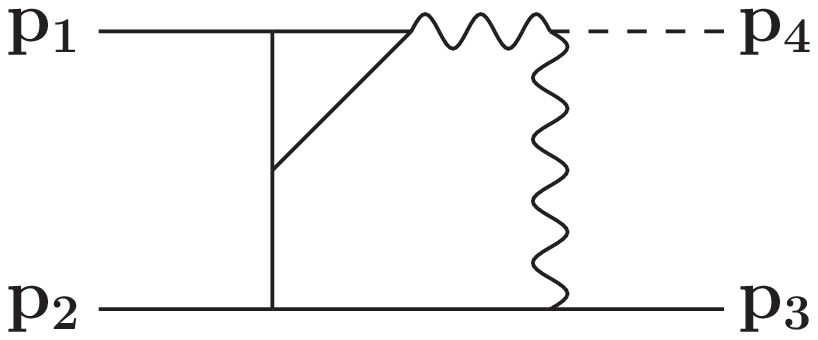}}\label{fig:T4}}
	\\
	\subfloat[T5: $\FIPL{0,1,1,1,1,1,1,1,0}$]{{\includegraphics[width=0.30\textwidth]{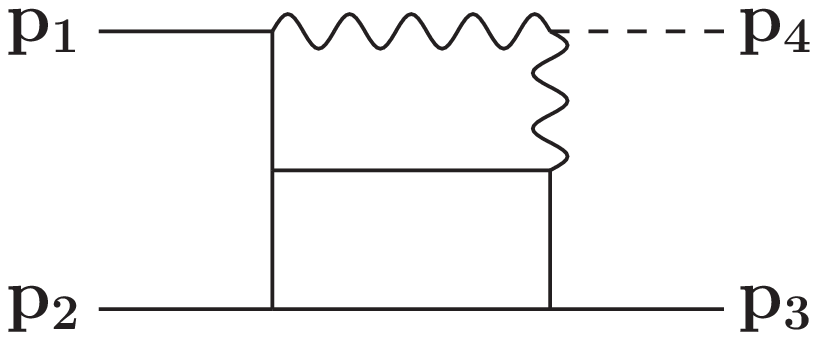}}\label{fig:T5}}
	\qquad
	\subfloat[T6: $\FIPL{1,1,1,1,1,1,1,0,0}$]{{\includegraphics[width=0.30\textwidth]{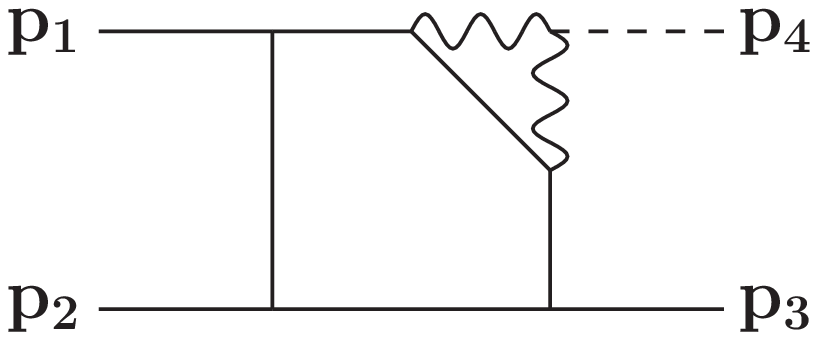}}\label{fig:T6}}	
	\caption{The six top sectors appearing in the amplitude. Straight (wavy) lines denote massless (massive) propagators. The dashed line indicates the Higgs boson. All momenta are taken to be incoming.}
    \label{topolo}
\end{figure}
\noindent With the definitions given in \autoref{tab:topos}, the top sectors of each integral family 
appearing in the amplitude are depicted in \autoref{topolo}. 
All the other diagrams contributing to the process can be obtained by permuting the external 
legs or by pinching the internal propagators.

We reduce the set of scalar integrals appearing in the amplitude to a basis of master integrals (\cite{Tkachov:1981wb,Chetyrkin:1981qh,Laporta:2001dd}) by first using \texttt{Reduze2}~\cite{Studerus:2009ye,vonManteuffel:2012np} to map the Feynman diagrams to the relevant integral families and then performing the reduction to master integrals with \texttt{KIRA}~\cite{Maierhoefer:2017hyi}.
We find that the independent helicity amplitudes can be written in terms of $69$ master integrals (modulus permutations of the external legs), see \autoref{sec:masters} for the full list. The master integrals obtained from the top sectors $T_2$ and $T_6$ have already been addressed in \cite{Bonetti:2020hqh}, while the remaining ones are new. We choose candidates for these new master integrals following the same prescriptions as for the $gg \to Hg$ case (for a detailed description, see \cite{Bonetti:2020hqh}).


\section{Computational details}
\label{sec:compt}

The computation of the amplitude is in principle straightforward: 
by following the approach discussed in \cite{Bonetti:2020hqh} for the $gg \to Hg$ amplitude, we compute the master integrals by direct integration over their Feynman-Schwinger parametrisation and we insert their expressions into the amplitude and collect common terms. However, in this case the size and complexity of both the expressions of the master integrals and the expressions of the form factors, together with the presence of UV and IR poles in $\epsilon$, makes the computation highly non-trivial.

\subsection{Evaluation of the master integrals}

We write the integrals of 
each integral family in \eqref{eq:FI} as integrals over Feynman-Schwinger parameters $x_i$ 
depending on the denominators $D_i$ from \autoref{tab:topos}.
In particular, the integrals are defined in terms of two polynomials $\U$ and $\F$ 
which are associated to each contributing top sector of the integral families considered \cite{Smirnov:AnalyticToolsForFeynmanIntegrals}.

Let us start considering the  integrals associated to the top-sector $T_3$.
In this case, the two polynomials read
\begin{align}
	\U &= x_3(x_2+x_4+x_5+x_7+x_8+x_9)+(x_4+x_5)(x_2+x_7+x_8+x_9)   \\
	\F &= m_V^2(x_2+x_7) \,\U - m_h^2 x_2 x_7(x_3+x_4+x_5)          \,,\nonumber\\
	   &\quad -s x_2(x_3 x_5+x_3 x_9+x_4 x_9 +x_5 x_9)-u x_7(x_3 x_4+x_3 x_8+x_4 x_8+x_5 x_8)   \,.
\end{align}
Applying the polynomial reduction algorithm \cite{Brown:PeriodsFeynmanIntegrals,Brown:2008um} as implemented in {\HyperInt} \cite{Panzer:2014caa}, these polynomials are found to be linearly reducible. This means in particular that, for a suitable order of the Schwinger parameters $x_i$, the successive integrals over each $x_i$ can be performed using {\HyperInt}. For this particular case, it is easy to verify that 
we can integrate in the order $x_4,x_5,x_3,x_8,x_9,x_2$.\footnote{The last variable, $x_7$, is not integrated over, but set to $x_7=1$.} 
In turn, this means that all integrals associated to these two polynomials can then be expressed as linear combinations $\sum_k R_k G_k$ of hyperlogarithms $G_k$, which are iterated integrals
\begin{align}\label{eq:hyper}
	\hlog{z}{\sigma_1,\ldots,\sigma_n} = \begin{cases}
		(\log z)^n/n! & \text{if $\sigma_1=\cdots=\sigma_n=0$ and} \\
		\int_0^z \frac{\td t}{t-\sigma_1} \hlog{t}{\sigma_2,\ldots,\sigma_n} & \text{otherwise.}
	\end{cases}
\end{align}
These functions are often called \emph{generalised harmonic polylogarithms} or \emph{multiple polylogarithms} in the physics literature, see for example Refs.~\cite{Remiddi:1999ew,Goncharov:1998kja,Goncharov:2001iea,Vollinga:2004sn}. 
Their arguments $\sigma_i,z$ and coefficients $R_k$ are algebraic functions of the variables $s,t,u,m_V^2$ (recall that $m_h^2=s+t+u$ is not independent). More precisely, the polynomial reduction shows that $\sigma_i,z,R_k$ are \emph{rational} functions in $s,t,u,m_V^2,r$ with the square root
\begin{align}
	r   = m_h^2 \sqrt{1-4 m_V^2/m_h^2}  \,.
\end{align}
This root arises in the last integration (over $x_2$). 
The analysis of the planar top sector $T_6$ is very similar and discussed in detail in \cite{Bonetti:2020hqh}. The conclusions are the same as for $T_3$.

The planar top sectors $T_4$ and $T_5$ are more complicated and introduce an additional square root 
in the final integration. This additional root is
\begin{align}\begin{aligned}
	r_{ust} &= \sqrt{s^2 u^2 +2su(t-s)m_V^2+(s+t)^2m_V^4}\quad\text{for $T_4$,}\\
	r_{sut} &= \sqrt{s^2 u^2+2su(t-u)m_V^2+(t+u)^2m_V^4}    \quad\text{for $T_5$.}
\end{aligned}\end{align}
The Symanzik polynomials of the topology $T_4$ are
\begin{align}
	\U &= x_3(x_1+x_2+x_4+x_7+x_8+x_9) +(x_1+x_4)(x_2+x_7+x_8+x_9)      \\
	\F &= m_V^2 (x_2 + x_7) \,\U -m_h^2 x_7(x_1x_2+x_1x_3+x_2x_3+x_2x_4)  \,,\nonumber\\
	   &\quad -sx_9(x_1x_2+x_1x_3+x_2x_3+x_2x_4)-ux_7(x_1x_8+x_3x_4+x_3x_8+x_4x_8)  \,.
\end{align}
These are linearly reducible and we can integrate e.g.\ in the order $x_1,x_4,x_3,x_8,x_9,x_2$ (setting $x_7=1$). The last integration over $x_2$ requires taking the roots $r$ and $r_{ust}$, therefore any integral of the topology $T_4$ can be written with prefactors $R_k$ and arguments of the hyperlogarithms $G_k$ being rational functions of $s,t,u,m_V^2,r,$ and $r_{ust}$.

The analysis of top sector $T_5$ is very similar. 
The conclusion is that all its integrals can be written in such a way that $R_k$ and the arguments of $G_k$ are rational in $s,t,u,m_V^2,r,$ and $r_{sut}$.

The non-planar topology $T_2$ was already discussed in \cite{Bonetti:2020hqh}. It produces prefactors $R_k$ and arguments in the hyperlogarithms $G_k$ that are rational functions of $s,t,u,m_V^2$ and four roots $r,r_t,r_u,r_{tu}$. We defined $r$ above, and the additional roots are
\begin{align}
	r_t = \sqrt{r^2-4 m_V^2 su/t}       \,,\qquad
	r_u = \sqrt{r^2-4 m_V^2 st/u}       \,,\qquad
	r_{tu} = \sqrt{1-4 m_V^2 /(t+u)}    \,.
\end{align}

Finally, also the non-planar topology $T_1$ turns out to be linearly reducible. The Symanzik polynomials are
\begin{align}
	\U &= (x_1+x_3)(x_2+x_7)+(x_1+x_2+x_3+x_7)(x_4+x_6+x_8) \,,\\
	\F &= m_V^2 (x_4+x_6) \,\U -m_h^2 x_6(x_1x_4+x_1x_7+x_2x_4+x_3x_4+x_4x_7) -s x_1x_2x_6  \nonumber\\
	   &\quad  - ux_3x_6x_7-t(x_1x_8(x_4+x_7)+x_2x_4(x_3+x_8)+x_4x_8(x_3+x_7))  \,,
\end{align}
and $x_1,x_3,x_2,x_7,x_8,x_4$ is a linearly reducible integration order (setting $x_6=1$).
The last integration (over $x_4$) introduces two additional square roots
\begin{align}
	r_{stu} &= \sqrt{s^2 t^2 + 2st(u-t)m_V^2+(t+u)^2m_V^4}  \,,\\ 
	r_{uts} &= \sqrt{u^2 t^2 + 2ut(s-t)m_V^2+(s+t)^2m_V^4} \,,
\end{align}
and, as a consequence, the integrals from topology $T_1$ will be expressed such that the prefactors $R_k$ and arguments of the hyperlogarithms $G_k$ are rational functions of $s,t,u,m_V^2,r,r_t$, $r_{stu},r_{uts}$.

\begin{table}
	\centering
	\begin{tabular}{ccccccccc}
	\toprule
	top sector    & $r$           & $r_t$         & $r_u$ & $r_{tu}$ & $r_{ust}$ & $r_{sut}$ & $r_{stu}$ & $r_{uts}$ \\
	\midrule
	$T_1$       &\textbullet    &\textbullet    & & & & &\textbullet  &\textbullet  \\
	$T_2$ &\textbullet  &\textbullet  &\textbullet  &\textbullet  & & & & \\
	$T_3$ &\textbullet  & & & & & & & \\
	$T_4$ &\textbullet  & & & &\textbullet  & & & \\
	$T_5$ &\textbullet  & & & & &\textbullet  & & \\
	$T_6$ &\textbullet  & & & & & & & \\
	\bottomrule
	\end{tabular}
	\caption{The entries in this table indicate that a root appears in the prefactors or hyperlogarithm arguments of some Feynman integral of the sector.}%
	\label{tab:root-topos}%
\end{table}
To summarize: all six top sectors are linearly reducible and hence the $\epsilon$-expansion of all corresponding Feynman integrals can in principle be computed by integration over Schwinger parameters using hyperlogarithms. The resulting expressions are linear combinations of hyperlogarithms, with arguments and prefactors that are rational functions of $s,t,u,m_V^2$ and a number of square roots as indicated in \autoref{tab:root-topos}.

Since master integrals with external legs crossed are also required in the computation of the form factors, additional roots appear, all obtained from permuting $s$, $t$, and $u$ in the definitions above. In particular, our results presented in the next section involve, in addition to the above, also the root
\begin{align}
    \begin{aligned}
        r_{tus} &= \sqrt{t^2 u^2 + 2tu(s-u)m_V^2+(s+u)^2 m_V^4} .
    \end{aligned}
\end{align}

\subsection{Construction of the form factors}

The analytic expressions of the master integrals expanded in $\epsilon \sim 0$ are inserted in the amplitude, obtaining results up to the finite part in $\epsilon$. We find that the form factors $F_{j}^{\textup{closed}}$ are finite in $\epsilon$, as expected. In fact, due to the presence of a closed quark loop coupled to the electroweak bosons, they are proportional to an electroweak coupling factor that is not present at LO. On the contrary, the form factors $F_j^{\textup{open}}$ contains UV and IR poles, up to order $1/\epsilon^2$ (see \autoref{sec:Cat} for a discussion on the poles and the extraction of the finite remainder for the amplitude).

The expressions involved in the amplitude are rather large, of the order of hundreds of megabytes, therefore special care must be invested in simplifying these expressions. 
Since the structure of the poles can be reconstructed starting form LO results thanks to the 
universal structure of IR singularities, we will focus here on polishing the necessary LO terms, 
namely the coefficient of $\epsilon^0$, $\epsilon^1$, and $\epsilon^2$, and only the $\epsilon$-finite part of the NLO ones.

First of all, we rescale all dimensionful variables in the form factors by $m_V^2$, so to obtain
\begin{align}
    F_{j,\epsilon^0}^{\textup{class}}(s,t,u,m_h^2,m_V^2) = \left( m_V^2 \right)^{-2} \tilde{F}_{j}^{\textup{class}}(T,U,\omega)  \,,
\end{align}
where $s=m_h^2-t-u$, $T=t/m_V^2$, $U=u/m_V^2$, $\omega=m_h^2/m_V^2$, and the overall $m_V^2$ factor is obtained by dimensional analysis.

In order to produce manageable expressions we scan each form factor and write it as a sum of algebraic prefactors (expressions containing ratios of polynomials possibly containing square roots) multiplied by transcendental expressions (containing hyperlogarithms, $\zeta$-functions, and $\mathrm{\pi}$)
\begin{align}
    \tilde{F}_{j}(T,U,\omega) = \sum_k R_{jk}(T,U,\omega) H_{jk}(T,U,\omega)	\,,
\end{align}
where we dropped the ${\textup{class}}$ index for shortness. We find that the simplest part of the amplitude is the LO one ($\tilde{F}^{\textup{open},(1)}_{\epsilon^0}$, $\tilde{F}^{\textup{open},(1)}_{\epsilon^1}$, $\tilde{F}^{\textup{open},(1)}_{\epsilon^2}$), followed by $\tilde{F}_{j}^{\textup{closed}}$, $\tilde{F}_{j}^{N_c}$, and finally $\tilde{F}_{j}^{1/N_c}$ as the most complex object of all.

We start by working on the rational prefactors $R_{jk}$. We apply the functions contained in the \texttt{Mathematica} \cite{MathematicaProg} package \texttt{MultivariateApart} \cite{Heller:2021qkz} together with their implementation in the computer language \texttt{Singular} \cite{DGPS} to each of the rational prefactors separately, 
in order to write them as linear combinations of rational expressions $M_h$ without spurious denominators
(we name such expressions \emph{rational monomials} from now on)
\begin{align}
	\label{red1}
	R_{jk}(T,U,\omega) = \sum_h a_{jkh} M_h(T,U,\omega)	\,.
\end{align}
Once the rational prefactors have been decomposed, we follow the
ideas described~\cite{Agarwal:2021grm} and look for linear relations among them using the rational monomials $M_j$ as independent variables. We stress that in this way we are \emph{not} guaranteed to find all relations
among the rational functions, since we are not
requesting
the partial fraction decomposition to be  unique across all rational prefactors. 
Nevertheless, we find that this procedure is already enough to reduce by about one half the number of independent rational prefactors, see \autoref{NumPref} for a comparison of the size of the form factors at different stages of simplification. 
Importantly, it has the advantage of not requiring the expensive computation of a Gr\"{o}bner basis across the full set of denominators that appear in the problem.

As a second step, we look for linear relations among all the rational monomials $M_h$ appearing in each individual rescaled form factor $\tilde{F}$ in order to find a basis of (linearly independent) elements $\overline{M}_h$ for the rational prefactors, obtaining the decomposition
\begin{align}
	R_{jk}(T,U,\omega) = \sum_h \overline{a}_{jkh} \overline{M}_h(T,U,\omega)	\,.
\end{align}

As final step, we search again for relations among different $R_{jk}$ expressed in terms of $\overline{M}_h$, obtaining a basis of rational prefactors.
All manipulations are fully analytical and we explicitly check that no more relations arise before moving to a subsequent step in the simplification procedure.

Clearly, we could have in principle omitted the step leading to Eq.~\eqref{red1}. 
As already stated above, the main reason for this preliminary reduction is that the majority of relations among the 
$R_{jk}$ are already found without working in the basis $\overline{M}_h$, 
reducing the number of different $M_h$ to analyze. This is a useful advantage, since  
the search for linear relations among the $M_h$
is extremely time consuming.

After having simplified the rational prefactors, we move to the combinations of transcendental functions
that multiply them. We inspect each term in the sum to check whether it is zero or not. We do it numerically (for different points in the Euclidean or Minkowski region) using the \texttt{Mathematica}~\cite{MathematicaProg} package \texttt{PolyLogTools}~\cite{Duhr:2019tlz}, and then discard null terms. We then again use these null relations to express hyperlogarithms of higher weight in terms of simpler functions, and we substitute them into the non-zero terms of the amplitude.
\begin{remark}
    Since we did not use a basis of master integrals with uniform transcendental weight, 
    our initial expressions for the form factors $\tilde{F}_j^{N_c}$ and $\tilde{F}_j^{1/N_c}$ had some contributions with hyperlogarithms of weight 5, which means $n=5$ in Eq.~\eqref{eq:hyper}. Those cancelled after the described simplification of the rational prefactors, leaving only hyperlogarithms of weights $\leq 4$. In contrast, the form factors $\tilde{F}_j^{\textup{closed}}$ had weight $\leq 4$ from the outset.
\end{remark}

\begin{table}[t]
	\centering
    \small
	\begin{tabular}{cccccc}
\toprule
~       &Original	&{Partial reduction} &{Monomial reduction}  &{Basis}		&{No zeroes}	\\
\midrule
$\tilde{F}_{1,\epsilon^0}^{\textup{open},(1)}$ &8~(31)  &7~(31)  &7~(24)  &6~(24)  &6~(24)  \\
$\tilde{F}_{2,\epsilon^0}^{\textup{open},(1)}$ &8~(33)  &7~(33)  &7~(30)  &6~(30)  &5~(24)  \\
$\tilde{F}_{1,\epsilon^1}^{\textup{open},(1)}$ &23~(70) &18~(70) &18~(50) &12~(50) &9~(21)	 \\
$\tilde{F}_{2,\epsilon^1}^{\textup{open},(1)}$ &22~(58) &16~(58) &16~(43) &12~(43) &9~(22)	 \\
$\tilde{F}_{1,\epsilon^2}^{\textup{open},(1)}$ &45~(87) &26~(87) &26~(65) &17~(65) &12~(37) \\
$\tilde{F}_{2,\epsilon^2}^{\textup{open},(1)}$ &44~(73) &23~(73) &23~(58) &17~(58) &10~(22) \\
\midrule
$\tilde{E}_{1}$ &46~(45)	&22~(45) &22~(28) &15~(28) &10~(17)	     \\
$\tilde{E}_{2}$ &46~(45) &22~(45) &22~(28) &15~(28) &10~(17)      \\
$\tilde{F}_{1,\epsilon^0}^{N_c}$		     &1410~(1454) &234~(1294)  &234~(1093)  &134~(1093) &100~(983)  \\
$\tilde{F}_{2,\epsilon^0}^{N_c}$		     &1413~(1389) &213~(1285)  &213~(1186)  &134~(1186) &117~(1169) \\
$\tilde{F}_{1,\epsilon^0}^{1/N_c}$	         &5526~(6789) &1174~(6788) &1100~(5177) &690~(3823) &325~(983)  \\
$\tilde{F}_{2,\epsilon^0}^{1/N_c}$	         &5524~(5905) &1139~(5894) &1139~(4604) &784~(4517) &460~(1169) \\
\midrule
$\tilde{B}_{1}$	  &4~(7)    &4~(7)	  &4~(7)    &4~(7)	  &~4(7)    \\
$\tilde{B}_{2}$	  &4~(7)	&4~(7)    &4~(7)    &4~(7)	  &~4(7)    \\
$\tilde{C}_{1}$   &43~(104) &35~(104) &32~(98)	&30~(98)  &30~(95)  \\
$\tilde{C}_{2}$   &41~(188) &34~(188) &33~(184)	&30~(184) &30~(123) \\
$\tilde{D}_{1}$   &67~(161) &64~(161) &61~(151) &54~(151) &54~(151)      \\
$\tilde{D}_{2}$   &93~(601) &91~(601) &89~(513) &54~(346) &54~(136)      \\

\bottomrule
	\end{tabular}
	\caption{Number of linearly-independent rational prefactors (and rational monomials) at different stages of the reduction procedure. We list the first three non-zero orders $\tilde{A}$ in the $\epsilon$ expansion for the LO amplitude, followed by the different component of the two-loop NLO amplitude and then by the different parts of the two-loop finite remainder, without taking into account the part proportional to $\log(m_V^2/\mu_R^2)$ (see \autoref{sec:Cat}).}
	\label{NumPref}
\end{table}

To conclude, we check numerically for linear relations among the $H_k$, using the PSLQ algorithm implemented in \texttt{Mathematica}. Once all linear relations among the $H_k$ have been applied to the expressions, we check again for relations among the new rational prefactors, to confirm that the expressions cannot be further reduced by means of the techniques described in this Section.\footnote{Except for some new zero rational prefactors in $F_2^{\textup{open},1/N_c}$, no further simplifications happen at this point.} 
We stress here that the surviving transcendental expressions are still not in a fully simplified form, 
and we expect that they can be further 
optimized both in size and for numerical evaluation by applying the same methods described in \cite{Bonetti:2020hqh}. 
This step is very elaborate due to the large number of different square roots, but it could
lead to substantially simpler results. Given its complexity, we leave this analysis for future work.

The expressions for the rescaled form factors $\tilde{F}_i$ presented here (and hence, for the form factors $F_i$) are valid in all of the three physical regions
\begin{align}
    \label{numericalphysics}
    \begin{aligned}
        0<m_h^2<4 m_V^2 \,,\quad\quad s>m_h^2 \quad\textup{and}\quad t,u<0 \qquad\qquad\textup{for $q \overline{q} \to H g$},\\
        0<m_h^2<4 m_V^2 \,,\quad\quad t>m_h^2 \quad\textup{and}\quad s,u<0 \qquad\qquad\textup{for $g \overline{q} \to H q$},\\
        0<m_h^2<4 m_V^2 \,,\quad\quad u>m_h^2 \quad\textup{and}\quad s,t<0 \qquad\qquad\textup{for $g q \to H \overline{q}$},
    \end{aligned}
\end{align}
by giving to $T$, $U$, and $\omega-T-U$ a small positive imaginary part for analytic continuation.

We kept both of the form factors separately through the simplification process in order to cross check our results thanks to the 
equivalence between them under the exchange $t \leftrightarrow u$, see \eqref{swap1} and \eqref{swap2}. For size reasons we provide only 
the first form factor for each component, the second being retrieved by the same exchange.


\section{Ultraviolet renormalization and infrared structure}
\label{sec:Cat}

The amplitude contains both ultraviolet and infrared singularities. While UV poles are removed through renormalization, IR singularities cancel only in infrared safe observables.

We remove the UV poles by renormalization of the strong coupling constant $\alpha_S$ in the $\overline{\textup{MS}}$ scheme. The NLO renormalization of the strong coupling constant reads \cite{Catani:1998bh}
\begin{align}
	\alpha_S = \overline{\alpha}_S\left(\mu_R^2\right) S_\epsilon^{-1} \left(\frac{\mu_R^2}{\mu_0^2}\right)^\epsilon \left[1 - \frac{\overline{\alpha}_S\left(\mu_R^2\right)}{2\mathrm{\pi}}\frac{\beta_0}{\epsilon}\right] + O\left(\overline{\alpha}_S^3\left(\mu_R^2\right)\right)	,
\end{align}
where $\alpha_S$ indicates the bare strong coupling constant, $\overline{\alpha}_S\left(\mu_R^2\right)$ the renormalized one, $\mu_R^2$ is the renormalization scale, $\mu_0^2$ is the dimensional regularization scale,
$S_\epsilon = (4\mathrm{\pi})^\epsilon \,\mathrm{e}^{\,- \epsilon\gamma}$, 
with $\gamma$ being the Euler--Mascheroni constant, and
\begin{align}
	\beta_0 = \frac{11 C_A - 4 N_f T_R}{6}	\,,
\end{align}
with $C_A = N_c$, $T_R = 1/2$, and $N_f$ the number of active flavors. From now on, we fix $\mu_0 = \mu_R$ for definiteness. 

We provide our results for the $\overline{\textup{MS}}$-renormalized form factors $\mathcal{F}_j^{\textup{class}}$ as a series expansion
in the renormalized strong coupling constant as follows
\begin{align}
    \begin{aligned}
        \mathcal{F}_{j}^{\textup{open}} &= S_\epsilon\sqrt{\frac{\overline{\alpha}_S\left(\mu_R^2\right)}{S_\epsilon}} \left[\mathcal{F}_{j}^{\textup{open},(1)} + \left( \frac{\overline{\alpha}_S\left(\mu_R^2\right)}{2 \mathrm{\pi}} \right) \mathcal{F}_{j}^{\textup{open},(2)} + \mathcal{O}\left( \overline{\alpha}_S^2 \right)\right] , \qquad \\
        \mathcal{F}_{j}^{\textup{closed}} &= S_\epsilon\sqrt{\frac{\overline{\alpha}_S\left(\mu_R^2\right)}{S_\epsilon}} \left[
        \left( \frac{\overline{\alpha}_S\left(\mu_R^2\right)}{2 \mathrm{\pi}} \right)\mathcal{F}_{j}^{\textup{closed},(2)} + \mathcal{O}\left( \overline{\alpha}_S^2 \right)\right] ,
    \end{aligned} \label{Fjren}
\end{align}
where  the coefficients read
\begin{align}
    \begin{aligned}
        \mathcal{F}_j^{\textup{open},(1)} &= S_\epsilon^{-1} F_j^{\textup{open},(1)}  \,,\\
        \mathcal{F}_j^{\textup{open},(2)} &= S_\epsilon^{-2} F_j^{\textup{open},(2)} - S_\epsilon^{-1} \frac{\beta_0}{2\epsilon} F_j^{\textup{open},(1)}  \,, \\
        \mathcal{F}_j^{\textup{closed},(2)} &= S_\epsilon^{-2} F_j^{\textup{closed},(2)}  \,,
    \end{aligned}
\end{align}
and the dependence of the form factors on the vector boson mass $m_V$ is  left
implicit from now on for ease of notation.\footnote{We stress that the extra factor
of $S_\epsilon$ in Eqs.~\eqref{Fjren} is chosen to reabsorb the 
terms proportional to $\gamma$
and $\log{(4 \pi)}$ coming from the loop integration on the electroweak vector bosons.} 
After UV renomalisation, $\mathcal{F}^{\textup{open},(2)}$ still contains IR poles. 
The structure of such singularities for QCD virtual NLO corrections is well-known and is fully captured by Catani's operator $\mathbf{I}^{(1)}$ \cite{Catani:1998bh}. We define therefore the finite reminders as
\begin{align}
	\label{catani}
    \begin{aligned}
	    \mathcal{F}_j^{\textup{open},(2)} &= \mathbf{I}^{(1)} \mathcal{F}_j^{\textup{open},(1)} + \mathcal{F}^{\textup{open},(2)}_{j,\textup{fin}}	\,, \quad \mathcal{F}_j^{\textup{closed},(2)} =  \mathcal{F}^{\textup{closed},(2)}_{j,\textup{fin}}
    \end{aligned}
\end{align}
where Catani's operator $\mathbf{I}^{(1)}$ reads 
\begin{align}
    \begin{aligned}
	    \label{catani1}
		\mathbf{I}^{(1)} = -\frac{1}{2} \frac{\mathrm{e}^{\,\epsilon \gamma}}{\Gamma(1-\epsilon)}
		&\left\{
		\mathcal{V}^{\textup{sing}}_q(\epsilon) \frac{2 C_F - C_A}{C_F} \left( - \frac{\mu_R^2}{s} \right)^\epsilon +	\right.\\
		&\left. +\frac{1}{2} \left( \mathcal{V}^{\textup{sing}}_g(\epsilon) + \frac{C_A}{C_F}\mathcal{V}^{\textup{sing}}_q(\epsilon) \right) 
		\left[ \left( - \frac{\mu_R^2}{t}  \right)^\epsilon 
		+ \left( - \frac{\mu_R^2}{u}\right)^\epsilon \right]
		\right\}	,
	\end{aligned}
\end{align}
where the Mandelstam invariants carry an implicit small positive imaginary part, 
$C_F = (N_c^2-1)/(2 N_c)$, and
\begin{align}
	\mathcal{V}^{\textup{sing}}_q(\epsilon) &= \frac{C_F}{\epsilon^2} + \frac{3 C_F}{2 \epsilon}	\,,\\
	\mathcal{V}^{\textup{sing}}_g(\epsilon) &= \frac{C_A}{\epsilon^2} + \frac{11 C_A - 2 N_f}{6 \epsilon}	\,.
\end{align}

In order to construct the finite remainders $\mathcal{F}^{\textup{open},(2)}_{\textup{fin}}$ 
we also require the one-loop results, $\mathcal{F}^{\textup{open},(1)}_{\textup{fin}}$, up to order $\epsilon^2$. 
To compute it, we use the same projectors and techniques illustrated in \autoref{sec:notation} and we compute the one-loop 
master integrals by direct integration over Feynman--Schwinger parameters. 
We provide the set of propagators and the master integrals for the one-loop case in \autoref{sec:masters}.
We have checked  numerically that the $\epsilon$-pole structures of $\mathcal{F}_j^{\textup{open},(2)}$ and 
$\mathbf{I}^{(1)} \mathcal{F}_j^{\textup{open},(1)}$ agree, in different points in Euclidean and Minkowski regions, 
using \texttt{Ginac} \cite{Vollinga:2005pk} and \texttt{PolyLogTools} \cite{Duhr:2019tlz}.

After all these manipulations, the one-loop and
two-loop finite reminders can be decomposed as a polynomial in $N_c$ and $N_f$. 
To simplify the notation and present our results
we write
\begin{align}
    \label{FRdec}
\mathcal{F}_{j,\textup{fin}}^{\textup{open},(1)} =  A_j	\,, \quad
	\mathcal{F}_{j,\textup{fin}}^{\textup{open},(2)} = N_f B_j + N_c C_j + \frac{1}{N_c} D_j	\,,
	\quad 	\mathcal{F}_{j, \textup{fin}}^{\textup{closed},(2)} =  E_j	\,.
\end{align}
We apply the same simplification procedure illustrated in \autoref{sec:compt} 
to the finite remainder as well (see \autoref{NumPref} and \autoref{IterPref} for the details).
We provide the analytical expressions for the coefficients $A_1$, $B_1$, $C_1$, $D_1$, $E_1$ 
(which have transcendental weight up to four), keeping the scale $\mu_R^2$ general, as supplementary material of this paper.
As discussed in Eq.~\eqref{swap1} and \eqref{swap2}, the remaining form factor can be obtained by exchanging $t \leftrightarrow u$.

\begin{table}[t]
	\centering
    \small
	\begin{tabular}{cSSr}
\toprule
~                           &{Hyperlogarithms}  &{Weight}           &\multicolumn{1}{c}{Size} \\
\midrule
$\tilde{A}_{\epsilon^0}$    &    11 &\SIrange{1}{2}{}   &   $\SI{0.5}{}\,\,\,\mathrm{kiB}$ \\
$\tilde{A}_{\epsilon^1}$    &    82 &\SIrange{1}{3}{}   &   $\SI{3.1}{}\,\,\,\mathrm{kiB}$ \\
$\tilde{A}_{\epsilon^2}$    &   386 &\SIrange{1}{4}{}   &  $\SI{16.7}{}\,\,\,\mathrm{kiB}$ \\
\midrule
$\tilde{B}_{1}$             &    11 &\SIrange{2}{3}{}   &   $\SI{1.7}{}\,\,\,\mathrm{kiB}$ \\
$\tilde{C}_{1}$             &  6907 &\SIrange{1}{4}{}   &   $\SI{0.7}{}\,\mathrm{MiB}$ \\
$\tilde{D}_{1}$             & 16755 &\SIrange{1}{4}{}   &   $\SI{1.3}{}\,\mathrm{MiB}$ \\
$\tilde{E}_{1}$             &   292 &\SIrange{0}{4}{}   &  $\SI{16.6}{}\,\,\,\mathrm{kiB}$ \\
\bottomrule
	\end{tabular}
	\caption{The complexity of our final expressions for the LO, and NLO finite remainder. The second column shows the number of different hyperlogarithm functions that appear.}
	\label{IterPref}
\end{table}

For reference, in \autoref{numerics} we provide numerical values for the  coefficients listed in Eq.~\eqref{FRdec} evaluated at the points
\begin{align}
    \label{pointstable}
    \begin{aligned}
    \sisetup{retain-explicit-plus}
    P_{q\overline{q}} &=\left\{\right.s\to\SI[retain-explicit-plus]{+375.1}{\giga\electronvolt}\,,\quad &&t\to\SI{-100.0}{\giga\electronvolt}\,,\quad &&u\to\SI{-150.0}{\giga\electronvolt}\left.\right\} ,\\
    P_{qg}            &=\left\{\right.s\to\SI{-150.0}{\giga\electronvolt}\,,\quad &&t\to\SI[retain-explicit-plus]{+375.1}{\giga\electronvolt}\,,\quad &&u\to\SI{-100.0}{\giga\electronvolt}\left.\right\} ,\\
    P_{\overline{q}g} &=\left\{\right.s\to\SI{-100.0}{\giga\electronvolt}\,,\quad &&t\to\SI{-150.0}{\giga\electronvolt}\,,\quad &&u\to\SI[retain-explicit-plus]{+375.1}{\giga\electronvolt}\left.\right\} ,
    \end{aligned}
\end{align}
with $\mu_R = m_h = \SI{125.1}{\giga\electronvolt}$ and $m_V = m_W = \SI{80.4}{\giga\electronvolt}$.

\begin{table}[t]
	\centering
    \small
	\begin{tabular}{crrr}
\toprule
~       &\multicolumn{1}{c}{$P_{q\overline{q}}$}	&\multicolumn{1}{c}{$P_{qg}$} 	&\multicolumn{1}{c}{$P_{\overline{q}g}$} \\
\midrule
$\tilde{A}_{1,\epsilon^0}$  &$+0.33014631\phantom{+01.25345679\,\iu}$&$+0.07937750+0.56737915\,\iu$ &$-0.20133007 + 0.56797218\,\iu$ \\
$\tilde{A}_{1,\epsilon^1}$  &$+0.86304102\phantom{+01.25345679\,\iu}$&$-0.29185479 + 1.53102839\,\iu$ &$-1.22164117 + 0.87055442\,\iu$ \\
$\tilde{A}_{1,\epsilon^2}$  &$+1.28181266\phantom{+01.25345679\,\iu}$&$-1.04954624 + 1.96804187\,\iu$ &$-1.87982473 + 0.03486591\,\iu$ \\
\midrule
$\tilde{B}_{1}$	            &$-0.00116710\phantom{+01.25345679\,\iu}$&$+0.15432174 + 0.02054956\,\iu$ &$+0.12722625 + 0.11327291\,\iu$ \\
$\tilde{C}_{1}$ 	        &$+0.51514217 + 0.00193541\,\iu$           &$-0.26266335 - 0.57780014\,\iu$ &$-0.49525929 - 1.62795284\,\iu$ \\
$\tilde{D}_{1}$          	&$+0.96451329 + 0.74965721\,\iu$	         &$+1.25415918 + 0.18720716\,\iu$ &$-1.20060441 + 0.68012840\,\iu$ \\
$\tilde{E}_{1}$             &$-0.11645020 - 0.21841056\,\iu$	         &$+0.51765798 + 0.00588251\,\iu$ &$+0.85351666 + 0.01089854\,\iu$ \\
\bottomrule
	\end{tabular}
	\caption{Numerical values of the different parts of the LO and NLO form factors, evaluated at points defined in Eq.~\eqref{pointstable}.}
	\label{numerics}
\end{table}


\section{Conclusions}
\label{conc}

In this paper, we computed the helicity amplitudes for the two-loop mixed QCD-Electroweak corrections to Higgs plus ject production in the quark-initiate channels $q g$, $\overline{q} g$, and $q \overline{q}$, keeping full dependence on the Higgs and EW vector boson masses. 
These amplitudes are the last missing ingredient, together with the $ggHg$ expressions computed in \cite{Bonetti:2020hqh,Becchetti:2020wof}, to construct electroweak corrections to Higgs plus jet production at hadron colliders. Starting from the Feynman diagrams contributing to the process we extracted the form factors by means of projectors and subsequently build the independent helicity amplitudes as a linear combination of scalar two-loop Feynman integrals. We use IBP relations to write the form factors as linear combinations of master integrals, which we subsequently evaluate by direct integration over Feynman/Schwinger parameters using the public code \HyperInt, exploiting their property of linearly reducibility. We simplify the outcome by separating the algebraic prefactors from the transcendental functions and rewriting the first ones in terms of a basis of linearly independent prefactors, while the transcendental expressions are investigated numerically to remove any linearly-dependent or zero term by means of the PSLQ algorithm. The result is formally more compact and reduces the number of transcendental functions to be evaluated to obtain numerical results. As a byproduct we also obtained expressions for the leading order one-loop form factors up to order $\epsilon^2$, necessary to verify the UV-renormalized IR pole structure described by the Catani's operator.

The computation described here shares many formal similarities with the $ggHg$ case discussed in \cite{Bonetti:2020hqh}, although being much more challenging on a practical level. The main reason for this is the fact that we are considering a NLO computation with a nonzero $\epsilon$-pole structure, that requires the evaluation of a larger number of master integrals, with more complex topologies and up to higher order in the $\epsilon$ expansion. The complexity and size of the $\epsilon^0$ results required additional refinement, obtained through the reduction to a basis of rational prefactors and the implementation of linear relations among the transcendental expressions obtained through PLSQ.
While the results presented here 
are not optimised for fast numerical evaluation due to the complexity of the alphabet involved, they can be evaluated straightforwardly 
in all relevant kinematical regions by specifying a suitable imaginary part for all kinematical invariants. Moreover, they can be used as a starting point for the construction of a
optimised analytic expressions in the various kinematical regions, for example following the ideas
described in~\cite{Heller:2019gkq} and already used in for example in~\cite{Bonetti:2020hqh}. 
We leave this non-trivial step for a future investigation.

\section*{Acknowledgments}
MB wishes to thank the Institute for Theoretical Particle Physics (TTP) at KIT and all of its admins 
for having provided access to their computational facilities, Federico Buccioni for help using the \texttt{Mathematica} package \texttt{MultivariateApart}, and Robert Harlander for his constant support. MB is much indebited to Federico Buccioni for his kind and resourceful help with the \texttt{OpenLoops} code.

We are grateful to V. Smirnov for various insights and collaboration on a related project.

MB is supported by the Deutsche Forschungsgemeinschaft (DFG) under grant no.\ 396021762 - TRR 257.

EP is funded by Royal Society University Research Fellow grant {URF{\textbackslash}R1{\textbackslash}201473}.

LT was supported by the Royal Society through grant URF{\textbackslash}R1{\textbackslash}191125, by the Excellence Cluster ORIGINS funded by the Deutsche Forschungsgemeinschaft (DFG, German Research Foundation) under Germany’s Excellence Strategy - EXC-2094 - 390783311 and by the ERC Starting Grant 949279 HighPHun.

\appendix 


\section{The master integrals}
\label{sec:masters}

\subsection{LO amplitude and master integrals}

The LO amplitude consists of $3$ Feynman diagrams, whose sum can be expressed in terms of $5$ master integrals (modulus permutation of the external legs). We proceed in the same way as for the virtual NLO amplitude, see \autoref{sec:notation} and \autoref{sec:compt}. The propagators of the integral family $\LO$ are listed in \autoref{ifLO}. The Symanzik polynomials are
\begin{equation*}
    \U=x_1+x_2+x_3+x_4
    \quad\text{and}\quad
    \F=-m_h^2 x_1 x_2 - t x_2 x_4 - u x_1 x_3 + m_V^2 (x_1+x_2)\,\U \,.
\end{equation*}

\begin{table}[t]
	\small
	\centering
	\begin{tabular}{cl}
\toprule
Denominator	&Integral family $\LO$	\\
\midrule
$D_1$		&$k^2-m_V^2$				\\
$D_2$		&$(k-p_1-p_2-p_3)^2-m_V^2$	\\
$D_3$		&$(k-p_2-p_3)^2$			\\
$D_4$		&$(k-p_2)^2$				\\
\bottomrule
	\end{tabular}
	\caption{Definition of the LO integral family $\LO$. The loop momentum is denoted by $k$. $m_V$ indicates the mass of the vector boson. The prescription $+ \ieps$ is understood for each propagator and not written explicitly.}
	\label{ifLO}
\end{table}

As master integrals, we have chosen
\begin{align}
		\label{LO-mi}
		\small
		\begin{aligned}
			&\FILO{3,0,0,0}\,, & &\FILO{2,1,0,0}\,, & &\FILOs{2,0,2,0}\,,  & &\FILOs{2,2,1,0}\,, & & & &\FILOs{1,1,1,1}\,.
		\end{aligned}
\end{align}
The integrals with an upper index $(6)$ are evaluated in $d=6-2\epsilon$ dimensions (without any upper index: in $d=4-2\epsilon$).

\subsection{Virtual NLO master integrals}

We expressed the NLO virtual amplitude in terms of $93$ master integrals (modulo permutations of the external legs). In particular
\begin{itemize}
	\item the planar integral family $\PL$ has four maximal topologies ($T_3$--$T_6$ in \autoref{topolo}),
	\begin{align*}
	    &\FIPL{1,1,1,1,1,1,1,0,0}\,, & & \FIPL{0,1,1,1,1,1,1,1,0}\,, \\ &\FIPL{1,1,1,1,0,0,1,1,1}\,, & & \FIPL{0,1,1,1,1,0,1,1,1}\,,
	\end{align*}
	and contains $70$ master integrals
	\begin{align}
		\label{PL-mi}
		\small
		\begin{aligned}
			&\FIPLs{0,0,0,2,0,2,2,2,0}\,, & &\FIPL{0,0,1,2,0,0,2,0,0}\,, & &\FIPLs{0,0,2,0,0,2,1,2,0}\,, \\
			&\FIPLs{0,0,2,0,1,1,2,1,0}\,, & &\FIPLs{0,0,2,0,2,0,2,1,0}\,, & &\FIPL{0,0,2,2,0,0,1,0,0}\,, \\
			&\FIPLs{0,0,3,0,0,2,0,2,0}\,, & &\FIPL{0,1,1,0,1,1,1,0,0}\,, & &\FIPL{0,1,1,1,0,1,0,1,0}\,, \\
			&\FIPL{0,1,1,1,0,1,1,0,0}\,, & &\FIPLs{0,1,1,1,1,1,1,2,0}\,, & &\FIPLs{0,1,1,1,1,1,2,1,0}\,, \\
			&\FIPL{0,1,2,0,0,2,0,0,0}\,, & &\FIPLs{0,1,2,0,0,2,2,1,0}\,, & &\FIPL{0,1,2,0,1,0,1,0,0}\,, \\
			&\FIPL{0,1,2,0,2,0,0,0,0}\,, & &\FIPLs{0,1,2,0,2,0,2,1,0}\,, & &\FIPL{0,1,2,1,0,0,1,0,0}\,, \\
			&\FIPL{0,1,2,1,0,1,0,0,0}\,, & &\FIPLs{0,1,2,2,0,0,2,0,1}\,, & &\FIPL{0,2,0,2,0,1,0,0,0}\,, \\
			&\FIPL{0,2,0,2,0,1,1,0,0}\,, & &\FIPLs{0,2,0,2,0,2,2,1,0}\,, & &\FIPL{0,2,1,1,0,1,1,0,0}\,, \\
			&\FIPLs{0,2,1,1,1,1,1,0,0}\,, & &\FIPLs{0,2,1,1,1,1,1,1,0}\,, & &\FIPLs{0,2,1,1,1,1,2,0,0}\,, \\
			&\FIPL{0,2,2,0,0,1,0,0,0}\,, & &\FIPLs{0,2,2,0,0,2,0,1,0}\,, & &\FIPLs{0,2,2,0,0,2,0,1,1}\,, \\
			&\FIPL{0,2,2,0,1,0,0,0,0}\,, & &\FIPL{0,2,2,0,1,0,1,0,0}\,, & &\FIPLs{0,2,2,0,1,1,0,1,0}\,, \\
			&\FIPLs{0,2,2,0,1,1,1,1,0}\,, & &\FIPLs{0,2,2,0,2,0,1,1,0}\,, & &\FIPL{0,2,2,1,0,0,1,0,0}\,, \\
			&\FIPL{0,2,2,1,0,1,0,0,0}\,, & &\FIPLs{0,2,2,1,0,1,1,0,0}\,, & &\FIPLs{0,2,2,1,1,0,1,0,0}\,, \\
			&\FIPLs{0,2,2,2,0,0,1,0,1}\,, & &\FIPLs{0,3,1,1,1,1,0,0,0}\,, & &\FIPLs{0,3,1,1,1,1,1,0,0}\,, \\
			&\FIPLs{0,3,2,0,1,1,0,1,0}\,, & &\FIPLs{0,3,2,1,1,0,1,0,0}\,, & &\FIPLs{0,4,1,1,1,1,0,0,0}\,, \\
			&\FIPL{1,0,1,0,1,0,2,0,0}\,, & &\FIPLs{1,0,1,1,1,0,3,0,0}\,, & &\FIPLs{1,0,1,1,1,0,4,0,0}\,, \\
			&\FIPL{1,0,2,0,1,0,2,0,0}\,, & &\FIPLs{1,0,2,1,0,0,2,0,1}\,, & &\FIPLs{1,0,2,1,0,0,3,0,1}\,, \\
			&\FIPL{1,1,1,0,0,1,1,0,0}\,, & &\FIPL{1,1,1,0,1,0,1,0,0}\,, & &\FIPL{1,1,1,0,1,0,2,0,0}\,, \\
			&\FIPL{1,1,1,0,1,1,1,0,0}\,, & &\FIPL{1,1,1,1,0,0,1,0,0}\,, & &\FIPL{1,1,1,1,0,1,1,0,0}\,, \\
			&\FIPLs{1,1,1,1,1,0,2,0,0}\,, & &\FIPLs{1,1,1,1,1,0,3,0,0}\,, & &\FIPLs{1,1,2,0,1,0,2,0,0}\,, \\
			&\FIPLs{1,1,2,1,0,0,2,0,1}\,, & &\FIPL{1,2,1,0,0,0,0,0,0}\,, & &\FIPLs{1,2,1,1,1,0,2,0,0}\,, \\
			&\FIPLs{2,0,2,0,0,0,2,1,0}\,, & &\FIPLs{2,0,2,0,0,0,2,1,1}\,, & &\FIPL{2,1,1,0,0,0,2,0,0}\,, \\
			&\FIPLs{2,1,2,0,0,0,2,1,0}\,, & &\FIPLs{2,1,2,0,0,0,2,1,1}\,, & &\FIPL{2,2,0,0,1,0,0,0,0}\,, \\
			& & &\FIPL{2,2,0,0,1,0,1,0,0}\,; & &
		\end{aligned}
	\end{align}
	\item the integral family $\NA$ (with maximal topology $\FINA{1,1,1,1,1,1,1,0,0}$, $T_2$ in \autoref{topolo}) contains $18$ master integrals:
	\begin{align}
		\label{NA-mi}
		\small
		\begin{aligned}
			&\FINAs{0,1,1,0,1,3,1,0,0}\,, & &\FINAs{0,1,1,1,1,2,1,0,0}\,, & &\FINAs{0,1,1,2,1,1,1,0,0}\,, \\
			&\FINAs{0,1,1,2,1,2,1,0,0}\,, & &\FINAs{0,1,1,3,1,0,1,0,0}\,, & &\FINAs{0,1,1,4,1,0,1,0,0}\,, \\
			&\FINA{1,1,0,1,1,0,1,0,0}\,, & &\FINA{1,1,0,1,1,1,1,0,0}\,, & &\FINAs{1,1,1,0,0,3,1,0,0}\,, \\
			&\FINAs{1,1,1,0,0,4,1,0,0}\,, & &\FINAs{1,1,1,1,0,2,1,0,0}\,, & &\FINAs{1,1,1,1,1,2,1,0,0}\,, \\
			&\FINAs{1,1,1,1,1,3,1,0,0}\,, & &\FINAs{1,1,1,2,0,1,1,0,0}\,, & &\FINAs{1,1,1,2,0,2,1,0,0}\,, \\
			&\FINAs{1,1,1,2,1,1,1,0,0}\,, & &\FINAs{1,1,1,3,0,0,1,0,0}\,, & &\FINAs{1,1,1,3,1,1,1,0,0}\,;
		\end{aligned}
	\end{align}
	\item the integral family $\NB$ (with maximal topology $\FINB{1,1,1,1,0,1,1,1,0}$, $T_1$ in \autoref{topolo}) contains $5$ master integrals:
	\begin{align}
		\label{NB-mi}
		\small
		\begin{aligned}
			&\FINBs{1,1,1,0,0,3,1,1,0}\,, & &\FINBs{1,1,1,1,0,3,1,1,0}\,, & &\FINBs{1,1,1,2,0,2,1,1,0}\,, \\
			&\FINBs{1,1,1,3,0,0,1,1,0}\,, & & & &\FINBs{1,1,1,3,0,1,1,1,0}\,.
		\end{aligned}
	\end{align}
\end{itemize}
The integrals with an upper index $(6)$ are evaluated in $d=6-2\epsilon$ dimensions (without any upper index: in $d=4-2\epsilon$).

\bibliography{Biblio}

\end{document}